\documentclass{nature}
\usepackage{graphicx}
\usepackage{amsmath}
\usepackage{amssymb}
\usepackage{textcomp}
\usepackage{url}

\newcommand{\ket}[1]{\vert#1\rangle}

\title{Flipping quantum coins}

\author{Guido Berl\'in$^{1}$, Gilles Brassard$^1$, F\'elix Bussi\`eres$^{2,3}$, Nicolas Godbout$^2$, Joshua A. Slater$^3$ \& Wolfgang Tittel$^3$}

\begin{document}

\maketitle

\begin{affiliations}
\item D\'epartement d'informatique et de recherche op\'erationnelle, Universit\'e de Montr\'eal, C.P.~6128, Succ.~Centre-Ville, Montr\'eal, Qu\'ebec, H3C~3J7 Canada
 \item Laboratoire des fibres optiques, D\'epartement de g\'enie physique, \'Ecole Polytechnique de Montr\'eal, C.P.~6079, Succ.~Centre-ville, Montr\'eal, Qu\'ebec, H3C~3A7 Canada
\item Institute for Quantum Information Science and Department of Physics and Astronomy, University of Calgary, 2500 University Drive NW, Calgary, Alberta, T2N~1N4 Canada
\end{affiliations}

\begin{abstract}
Coin flipping is a cryptographic primitive in which two distrustful parties wish to generate a random bit in order to choose between two alternatives\cite{Blum81}. This task is impossible to realize when it relies solely on the asynchronous exchange of classical bits: one dishonest player has complete control over the final outcome. It is only when coin flipping is supplemented with quantum communication that this problem can be alleviated\cite{ATVY00,SR02a,SR02b,Ambainis04,CK09}, although partial bias remains\cite{LC98,MSC98}. Unfortunately, practical systems are subject to loss of quantum data\cite{BM04}, which restores complete or nearly complete bias in previous 
protocols\cite{ATVY00,SR02a,SR02b,Ambainis04,CK09,MVUZ05,NFPM08}. We report herein on the first implementation of a quantum coin-flipping protocol that is impervious to loss. Moreover, in the presence of unavoidable experimental noise, we propose to use this protocol sequentially to implement many coin flips, which guarantees that a cheater unwillingly reveals asymptotically, through an increased error rate, how many outcomes have been fixed. Hence, we demonstrate for the first time the possibility of flipping coins in a realistic setting. Flipping quantum coins thereby joins quantum key distribution\cite{BB84,GRTZ02} as one of the few currently practical applications of quantum communication. We anticipate our findings to be useful for various cryptographic protocols and other applications, such as an online casino, in which a possibly unlimited number of coin flips has to be performed and where each player is free to decide at any time whether to continue playing or not. 
\end{abstract}

Coin flipping is the art of tossing a coin to allow two parties to choose between two alternatives in the least biased way. The importance of this primitive led Manuel Blum to introduce ``coin flipping by telephone'', in which the two spatially separated parties do not necessarily trust each other but still wish to ensure that the outcome is unbiased\cite{Blum81}. Throughout this Letter, we only consider asynchronous protocols, which consist of a sequence of rounds in which Alice and Bob alternate in sending classical messages to each other. For any classical coin-flipping protocol, one of the parties has the possibility to deterministically choose the outcome, in which case we say that the protocol is broken. 

In the quantum world, this is no longer true\cite{ATVY00,SR02a,SR02b,Ambainis04,CK09}. While no protocol with zero bias can exist\cite{LC98,MSC98}, the probability for a cheater to fix the outcome can be asymptotically reduced to $1/\sqrt{2} \approx 70.7\%$\cite{CK09}. Thus, quantum communication allows one to implement a cryptographic primitive that is impossible using only classical communication.

The typical structure of most previous protocols is as follows. Alice sends a quantum state $\ket{\psi}$ to Bob, chosen from an agreed upon set, that conceals a bit~$a$. Bob returns a classical bit $b$. Alice then discloses which $\ket{\psi}$ was sent, thereby revealing~$a$. The outcome $c$ of the coin flip is the exclusive~OR of $a$ and $b$, denoted $c = a\oplus b$. For quantum to be better than classical coin flipping, Bob must not be able to determine the value of $a$ with certainty from a measurement of $\ket{\psi}$ before returning his bit. Furthermore, Alice must not be able to declare a value of $a$ that depends on the value of $b$ without risking being caught cheating through Bob's measurement of~$\ket{\psi}$. 

As usual in quantum communication, quantum states are encoded into photons, which are susceptible to loss in the transmission channel and measurement apparatus. If Bob's measurement happens after Alice revealed her bit, this allows him to cheat by pretending that the photon was lost whenever he is not happy with~$a$. Moreover, in any realistic implementation, which includes imperfect state preparation and noisy transmission channel and detectors, a cheating Bob can always pretend that his measurement did not confirm Alice's declared state $\ket{\psi}$. If Alice allows the protocol to be restarted until Bob declares that he is satisfied, the latter can always completely control the outcome\cite{BM04}. 

A protocol that is not broken in the presence of loss was recently obtained\cite{NFPM08}. The improvement is, however, marginal
as Alice can still choose the outcome with near certainty in a realistic setting. Moreover, the bias increases with loss. To be of practical use, a protocol should be \emph{loss~tolerant}, which we define as being impervious to loss. 

In this Letter, we present the first experimental demonstration of a loss-tolerant coin-flipping protocol along with the optimal cheating strategies for Alice and Bob. The complete theoretical description and analysis of this protocol, including the proofs for the optimality of the cheating strategies, is presented elsewhere\cite{BBBG09}. We also propose a new task for which this protocol remains useful in the presence of experimental noise. Hence, we demonstrate for the first time the possibility of flipping coins in a realistic setting. Flipping quantum coins thereby joins quantum key distribution\cite{BB84,GRTZ02} as one of the few currently practical applications of quantum communication.

Let us first describe our protocol when both parties are honest. Alice sends a qubit whose state $\ket{\psi_{x,a}}$ is chosen randomly among the following, previously agreed upon set (see Fig.~1):
\begin{eqnarray} \label{eqn:states}
\begin{array}{ccc}
 \ket{\psi_{0,0}} &=& \ket{0} \\
    \ket{\psi_{0,1}} &=& \ket{1} \\
    \ket{\psi_{1,0}} &=& \ket{\varphi^+} \\
    \ket{\psi_{1,1}} &=& \ket{\varphi^-} 
\end{array}
\end{eqnarray}
where $\ket{\varphi^+} = \cos \varphi \ket{0} + \sin \varphi \ket{1}$, $\ket{\varphi^-} = \sin \varphi \ket{0} - \cos \varphi \ket{1}$ and $0^{\circ} < \varphi \le 45^{\circ}$. Here, $x$ and $a$ represent Alice's basis and bit, respectively. Bob then attempts to measure Alice's qubit in a basis $y \in \{0,1\}$ from the set described above, chosen at random. If Bob does not detect the qubit, he asks Alice to send another randomly selected state. If Bob detects the qubit, he sends a random bit $b$ to Alice. When Alice receives $b$, she reveals $x$ and $a$ to Bob. If $y=x$, Bob's measurement outcome should agree with Alice's declared state $\ket{\psi_{x,a}}$, in which case $a$ is accepted. In case of a disagreement, Bob declares a mismatch. If $y \ne x$, Bob has no way to verify Alice's claim with certainty and he must accept her bit on faith. The outcome of the protocol is $c = a \oplus b$.

The loss tolerance of our protocol stems from two features. The first is that Bob's measurement happens before Alice reveals her bit $a$. The second is that Bob gains no advantage in falsely declaring that Alice's qubit was lost. In particular, it is physically impossible for Bob to determine Alice's bit $a$ with certainty, given a single copy of $\ket{\psi_{x,a}}$. The performance of optimal cheating strategies depends on the value of $\varphi$. For the states given in Eq.~(\ref{eqn:states}) with $\varphi = 45^{\circ}$, which we refer to as the \emph{BB84} states (see Fig.~1-a), Alice's maximum probability to fix the outcome, $P_A$, is $(6+\sqrt{2})/8 \approx 92.7\%$; she is caught cheating with the complementary probability, i.e.~when a mismatch occurs. Bob's equivalent probability, $P_B$, is $(2+\sqrt{2})/4 \approx 85.4\%$, which makes the use of these symmetrically distributed states unfair as Alice can cheat better. By setting $\varphi = \arccos(4/5) \approx 36.9^{\circ}$, which results in what we call the \emph{fair} states (see Fig.~1-b), this asymmetry is removed, leading to $P_A = P_B = 90\%$. For both sets of states, Alice's optimal cheating strategy consists of randomly sending one of the two orthogonal states $\ket{A_0}$ and $\ket{A_1}$ that are positioned symmetrically between states representing different bit values $a$, as shown on Fig.~1. This choice allows her to always declare an $x$ and $a$ that will produce the outcome of her choice while minimizing her probability of being caught cheating. Bob's optimal cheating strategy consists of measuring the received qubit in basis $\{\ket{B_0}, \ket{B_1} \}$, where $\ket{B_i}$ is positioned symmetrically between the states that correspond to the bit value $a = i$, as shown on Fig~1. This maximizes his probability to guess the value of Alice's bit. Note that we do not consider the case in which both Alice and Bob are cheating, as the goal of the protocol is to protect the honest player only.

\begin{figure}
\caption{\textbf{Honest and cheating states of the loss-tolerant protocol.} The states used are represented on a great circle of the Bloch sphere. \textbf{a)} BB84 states, corresponding cheating states $\ket{A_0}$ and $\ket{A_1}$ for Alice, equal to $\ket{\varphi_{A}^+}$ and $\ket{\varphi_{A}^-}$ with $\varphi_{A} = 67.5^{\circ}$, and corresponding cheating states $\ket{B_0}$ and $\ket{B_1}$ for Bob, equal to $\ket{\varphi_{B}^+}$ and $\ket{\varphi_{B}^-}$ with $\varphi_{B} = 22.5^{\circ}$. \textbf{b)} Fair states and the corresponding cheating states defined as for the BB84 states but with $\varphi_{A} \approx 63.4^{\circ}$ and $\varphi_{B} \approx 18.4^{\circ}$.}
\end{figure}

The resulting protocol is impervious to loss. However, it is not tolerant to noise on a single coin flip. Indeed, in case of a mismatch, Bob cannot know with certainty if it is caused by a cheating Alice or by noise in the state preparation, transmission or detection. In fact, all known protocols for flipping a single coin are broken in the presence of noise\cite{BM04} and it remains an open question whether such a noise-tolerant protocol exists at all.  We propose considering a different task: Alice and Bob need to flip many coins, possibly an unlimited number, and have the possibility to stop playing whenever they want and for whatever reason, including that they no longer trust the other player. This scenario might apply to an online casino where the client and the house do not trust each other but may still wish to play indefinitely. According to this task, which we shall call \emph{sequential coin flipping}, each player can secretly decide, given sufficient statistics, the maximum tolerable error rate (the rate at which mismatches occur). Obviously, this threshold should be smaller than the rate induced solely by optimal cheating and larger than the intrinsic error rate of the experimental setup. The observed error rate asymptotically reveals how many coin flips the cheater was able to fix, a measure that is impossible to assess in a single coin flip. Note, however, that sequential coin flipping requires a loss-tolerant single coin-flipping protocol such as ours. Otherwise, the cheater would exploit losses to fix the outcome, as discussed above, instead of pursuing the strategy that involves falsely declaring a mismatch. We stress the difference between sequential coin flipping and random bit-string generation, in which Alice and Bob generate at once a random string of bits with predetermined length\cite{BM04}. Indeed, this string cannot be used for sequential coin flipping as the players know in advance all the bits of the sequence, hence they can decide when to stop depending on the outcome of future coin flips. Whereas random bit-string generation can be done classically\cite{Buhrman07}, we conjecture that sequential coin flipping cannot as it would be composed of many individual coin flips that can be broken by a potential cheater. 

For the experimental implementation of the protocol, the restrictions on Alice's qubit source are very stringent, even more than in the closely related primitive of quantum key distribution\cite{BB84,GRTZ02,Scarani08}, as Bob is an adversary who is potentially cheating. In particular, if Alice is using attenuated laser pulses, then Bob can easily break the protocol using his \textit{honest} measuring apparatus, as discussed in the Methods. With current technology, Alice's only practical choice is to use a source of entangled qubits. Projecting one qubit at Alice's remotely prepares a random state on the second qubit being sent to Bob\cite{BBM92}.

Our experimental setup is detailed in Fig.~2. Time-bin entangled photonic qubits\cite{BGTZ99} in the state
\begin{equation}
  \ket{\Phi^+} = \frac{1}{\sqrt{2}}(\ket{e}_A\ket{e}_B+ \ket{\ell}_A\ket{\ell}_B)
\end{equation}
are created, where $\ket{e}_{A(B)}$ and $\ket{\ell}_{A(B)}$ represent the \emph{early} and \emph{late} time-bin states of Alice (Bob) and are associated to the generic states $\ket{0}$ and $\ket{1}$ used above to describe the protocol. One qubit remains in Alice's laboratory where it is randomly projected on one of the four states defined in~(\ref{eqn:states}) using a novel free-space universal time-bin analyser (UTBA) (see the Methods for more details). This has the effect of remotely preparing the other qubit in the same state. The latter is sent to Bob over the fibre communication link, which consists of 10~m of polarization maintaining (PM) fibre. Note that the PM fibre can be replaced with standard fibre and polarization control\cite{MCMHT09}. Bob, by virtue of his fibre UTBA, then measures his time-bin qubit in a randomly chosen basis $y$. To the best of our knowledge, these UTBAs enable for the first time projection measurements of time-bin qubits in arbitrary bases, which facilitates the implementation of the fair protocol and of all the cheating strategies. We characterized the quality of our source by measuring entanglement visibilities of at least $(91.0 \pm 2.8)$\%.
\begin{figure}
\caption{\textbf{Experimental setup.} A laser diode sends 50~ps pulses at 530.6~nm wavelength through an interferometer with path length difference equivalent to 1.4~ns travel time difference. The pulses emerge in an even superposition of two well defined time-bins that we label the \textit{early} and \textit{late} bins and then propagate into a non-linear, periodically poled lithium niobate crystal (PPLN), thereby creating time-bin entangled qubits at 807 and 1546~nm wavelengths through spontaneous parametric downconversion. The two qubits are separated at the dichroic mirror (DM). As explained in the Methods, the free-space and fibre UTBAs randomly project Alice's and Bob's qubits onto one of the four states defined in~(\ref{eqn:states}), where $\varphi$ is selected by the orientation of the output half-wave plate (HWP) at Alice's and the polarization controller at Bob's.  The coincidence detections are monitored using a time-domain converter (TDC) and analysed in real time.}
\end{figure}

We performed sequential coin flipping either with the BB84 or the fair states, in both cases with honest players or one cheater, as determined by the settings of the UTBAs. Let us consider the honest cases first. We measured the error rate $P^*$ and the probabilities $P_0$ and $P_1$ of outcomes $c = 0$ and~$1$ per coincidence detection (which we define as a coin-flip \emph{instance}). As shown on Fig.~3-a, the $P^*$ obtained when using either the BB84 or the fair states is smaller than $2\%$, i.e.~less than the lowest error rate that a cheater would theoretically induce by cheating on every single coin flip ($\gtrsim7.3\%$ for the BB84 states and $10\%$ for the fair states). This ensures that cheating is noticeable over several coin flips. Furthermore, the outcome probabilities $P_0$ and $P_1$ are equal within one standard deviation, which demonstrates that the protocol does indeed result in unbiased bits in the honest case.

Next we consider the cases where either Alice or Bob tries to fix the outcome $c$ of every coin flip. The cheater's UTBA was aligned for optimal cheating. For each instance, we let the cheater choose a desired value for $c$, uniformly distributed. After many coin-flip instances, we measured the probability $P_A$ $(P_B)$ for Alice (Bob) to fix the outcome to the desired value, as well as $P^*$. We assumed that a cheating Bob would always declare a mismatch when he was unhappy with the outcome. 

As a first observation of the results presented on Fig.~3-b and 3-c, we note that the values obtained for $P_A$ and $P_B$ equal, within one standard deviation, the theoretical maximum minus the error rate $P^*$ measured in the honest case, which indicates that the UTBAs were well aligned for each case. Next, as shown on Fig.~3-b, we see that the BB84 states are clearly unfair as $P_A$ is higher than $P_B$. As a result, the error rate for a cheating Alice is less than for a cheating Bob. On Fig.~3-c, we see that this asymmetry is removed when using the fair states. Indeed, $P_A$ and $P_B$ are now equal within one standard deviation. Furthermore, when using the BB84 states, $P_A = (91.1 \pm 0.1)\%$, which is significantly higher than $90\%$, i.e.~we are able to show that Alice can indeed cheat better than what is theoretically possible with the fair states. Similarly, when using the fair states, $P_B = (88.4 \pm 0.1)\% > 85.4\%$, which demonstrates that Bob can cheat better than what is theoretically possible with the BB84 states. Finally, we note that the error rate $P^*$ increases from less than $2$\% to at least $8.9$\% when moving from the honest case to the case where one player always tries to cheat optimally. This clearly shows that the cheater is revealed over many coin-flip instances and the sequential coin-flipping protocol proposed here can be implemented with current technology.

\begin{figure}
\caption{\textbf{Experimental results of sequential coin flipping.}  The column plots show the \textbf{a)} honest player cases, \textbf{b)} one cheater cases with BB84 states, \textbf{c)} one cheater cases with fair states. All data collection runs consisted of at least 80 thousand coin-flip instances. The one-standard-deviation uncertainties on $P_0$, $P_1$, $P_A$ and $P_B$ are at most 0.15\%.  The uncertainty on $P^*$ is at most $0.04\%$ in the honest cases and at most $0.13\%$ in the cases where one player cheats.}
\end{figure}

\begin{methodssummary} 
The pump laser was operated at a repetition rate of 10~MHz (see Fig.~2). The 10~mm long PPLN crystal, with a $7.05$~\textmu m grating period, was heated to 176$^{\circ}$C. The mean number of photon pairs created per pump pulse was about 0.05, which set the probabilities to create one and two pairs to 4.8\% and 0.12\%, respectively\cite{BSGT08}. 

Detection events were acquired by a time to digital converter (TDC), which measures delays between a start signal and several stop signals. The detection signals from Alice's free-running Si-based single photon detectors were pre-processed with an electronic mixer (see Fig.~2). The \emph{trigger} signal was generated when a detection at either $S_{1}$ or $S_{2}$ occurred. It emerged synchronously with the laser clock (\textit{clk}). The signal was used to gate Bob's InGaAs-based single photon detectors during a 7~ns activation window. The signal \emph{ready}, which was emitted only when both detectors were ready to detect, was used to start the TDC\@. This ensures that the statistics were not biased by the dead-time of Bob's detectors. The detections at $I_1$ and $I_2$, as well as the signal $S_2 \wedge \textit{clk}$ and $S_1 \vee S_2$ served as stop signals, where $\wedge$ denotes the logical AND and $\vee$ the logical OR. This allowed us to register all possible coincidence detection events, where the detection slots, early, middle and late, were narrowed down to 400~ps. This particular event selection to retrieve information about detections at $S_{1}$ and $S_{2}$ was chosen due to hardware considerations in Alice's mixer.
\end{methodssummary}

\begin{methods}

\subsection{Requirements on Alice's source of qubits.}
The adversarial nature of the players forces Alice to consider side channels that a cheating Bob could exploit. For instance, when using attenuated laser pulses or a heralded single photon source\cite{Scarani08}, Alice will sometimes send multiple photons. In this case, all photons would be prepared in the same qubit state $\ket{\psi_{x,a}}$. In the presence of loss, this allows a cheating Bob to declare that the photon was lost unless he detects two photons in different bases using his \emph{honest} measuring apparatus. When this happens, Bob can conclusively determine Alice's bit $64\%$ of the time (with the fair states), in which case only will he declare the photon detected, thereby completely breaking the protocol\cite{BBBG09}. To prevent this, Alice could use a perfect source of single photons. However, this is not practical with current technology. A more realistic choice is to use a source of maximally entangled photonic qubits for which a projection at Alice's remotely prepares a state at Bob's\cite{BBM92}. Should the source emit multiple pairs of entangled qubits, the state of the qubits sent to Bob are uncorrelated, hence Bob cannot gain any information from multi-pair events. 

\subsection{Universal time-bin analysers}
The free-space universal time-bin analyser\cite{BSBLG06} \linebreak (UTBA) shown in Fig.~2 can be understood as follows (it will be described in details elsewhere, F.\,B., J.\,A.\,S., N.\,G., \&~W.\,T., in preparation). The polarization of the incident time-bin qubit is first rotated to 45$^{\circ}$ with respect to the linear polarization transmitted by the input polarizing beamsplitter (PBS). After passing through an interferometer with large path length difference, the qubit emerges in three chronologically ordered times slots separated by 1.4~ns that we label \emph{early}, \emph{middle} and \emph{late}. In the middle slot, the initial time-bin qubit is mapped on a polarization qubit, which can be analysed in any basis using standard waveplates, a PBS and detectors. This implements the detection in the $x=1$ basis at Alice's, where the angle $\varphi$ is determined by the orientation of the half-wave plate (HWP) located at the output of the interferometer. Detection in the early (late) slots corresponds to a projection on $\ket{e}_{A}$ $(\ket{\ell}_{A})$, and this implements a measurement in the $x=0$ basis. Therefore, the detection time at the single photon detectors $S_{1}$ and $S_{2}$ determines Alice's basis. This is similar to previous projection measurement schemes for time-bin qubits\cite{TBZG00}, yet without the restriction to mutually unbiased bases. 

Bob's fibre UTBA is the fibre optics equivalent of the free-space version. The input fibre, as well as the two arms of the interferometer, are made of PM fibre. The output of the interferometer is a standard fibre and the angle $\varphi$ is selected by a polarization controller. Here again, the detection time at the single photon detectors $I_{1}$ and $I_{2}$ determines Bob's basis.

\subsection{Entanglement visibility and phase alignment}
To measure the entanglement visibility, we first set the free-space UTBA to project on the states $\ket{e} \pm \textrm{e}^{i \alpha}\ket{\ell}$ and the fibre UTBA to project on $\ket{e} \pm \textrm{e}^{i \beta}\ket{\ell}$, where $\alpha$ and $\beta$ are unknown phases (we omitted normalization). Then, for a given detection at one of Alice's detectors $S_i$, the probability $p_{ij}$ for a coincident detection in one of Bob's detectors $I_j$ was measured while the phase $\alpha$ of the free-space UTBA was scanned by displacing the piezo-actuated retroreflector of the long arm. This caused the usual sinusoidal variation of $p_{ij}$ from which visibilities $\ge (91.0 \pm 2.8)\%$ were measured. We adjusted $\alpha$ to maximize $p_{11}$, indicating that Alice's and Bob's UTBAs are preparing and measuring in the same basis, i.e.~$\alpha = \beta$. This alignment procedure preceded each data collection run.
\end{methods}

\begin{addendum}
\item[Acknowledgements]
G.\,Br.~is supported in part by Canada's Natural Sciences and Engineering Research Council ({\sc Nserc}), the Canada Research Chair program, the Canadian Institute for Advanced Research ({\sc Cifar}), Quantum\emph{Works} and the Institut transdisciplinaire d'informatique quantique ({\sc Intriq}). F.\,B.~is supported in part by the Fonds qu\'eb\'ecois de la recherche sur la nature et les technologies (\textsc{Fqrnt}), the Canadian Institute for Photonics Innovations ({\sc Cipi}) and an {\sc Nserc} Canada \mbox{Graduate} \mbox{Scholarship}. N.\,G.~is supported in part by the Centre d'optique, photonique et lasers (\textsc{Copl}), Quantum\emph{Works}, {\sc Nserc}, {\sc Cipi} and {\sc Intriq}. J.\,A.\,S.~is supported in part by the Alberta Ingenuity Fund (\textsc{Aif}) and the Information Circle of Research Excellence (i\textsc{core}). W.\,T.~is supported in part by \textsc{Nserc}, General Dynamics Canada (\textsc{Gdc}), i\textsc{core}, Quantum\emph{Works}, Alberta Advanced Education and Technology (\textsc{Aaet}) and the Canada Foundation for Innovation (\textsc{Cfi}).
 \item[Author Contributions] The theory was developed by G.\,Be., G.\,Br., F.\,B. and N.\,G. The experiment was performed by F.\,B., J.\,A.\,S. and W.\,T.
 \item[Author information] Correspondence and requests for materials should be addressed to F.\,B.~(felix.bussieres@polymtl.ca).
\end{addendum}

\AtEndDocument{
%\newpage
\noindent \ \\
Figure~1:\\ \ \\
\includegraphics[scale=1]{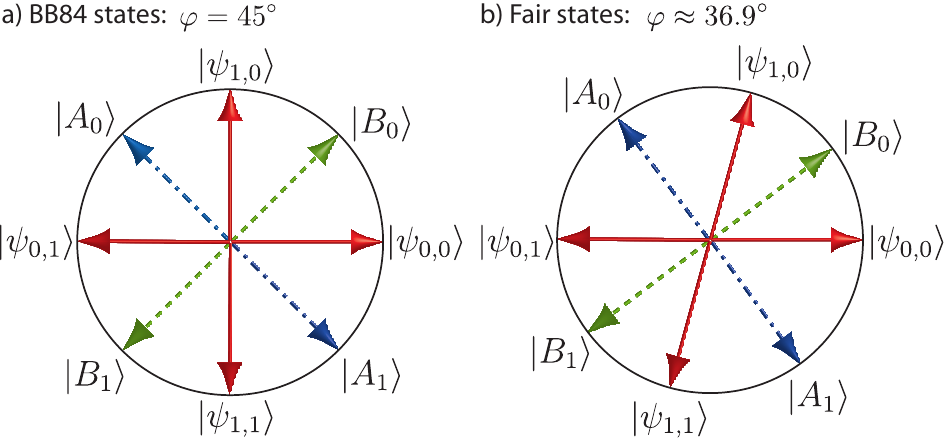}\\
\noindent
Figure~2:\\ \ \\
\includegraphics[scale=1]{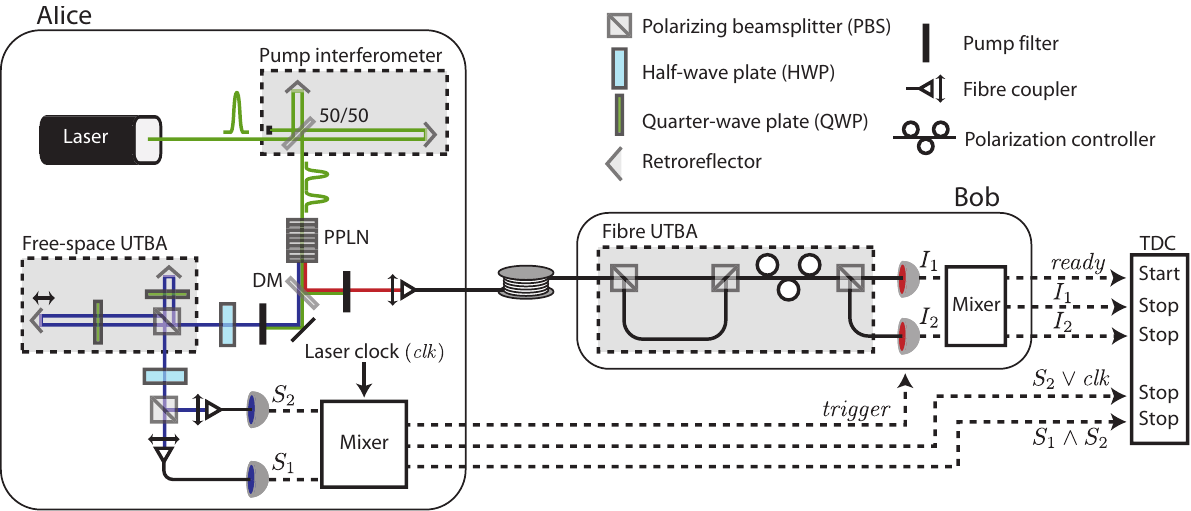} \\
\noindent Figure~3:\\ \ \\
\includegraphics[scale=1]{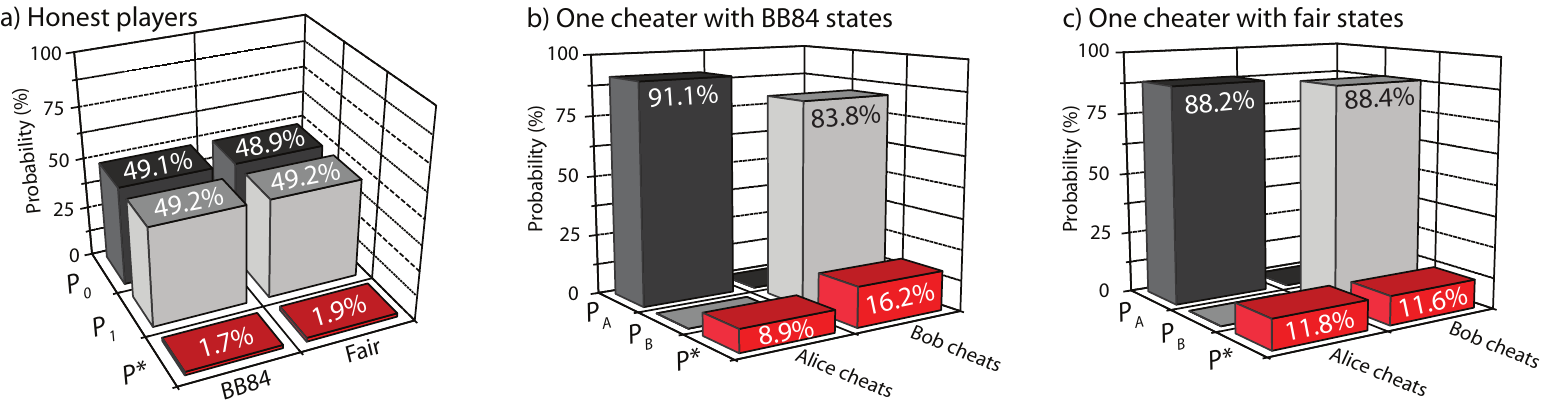}
}

\end{document}